\def\vday{{January 11,1993}}
\documentstyle[preprint,revtex]{aps}

\begin{document}
\draft
\centerline{\small{Extended Version: \vday}
            \hfill FERMILAB-PUB-92/172--T}
\centerline{\small{Original preprint in July 1992}
            \hfill NUHEP--TH--92--14}
\pagestyle{empty}
\begin{title}
CP Violation in the Decay of Neutral Higgs Boson  \\
into $t-{\bar t}$ and $W^+-W^-$
\end{title}
\author{
Darwin Chang$^{(1,2,3)}$,
Wai--Yee Keung$^{(4,2)}$\\
\quad\\  \quad \\
}
\begin{instit}
$^{(1)}$Department of Physics and Astronomy,
Northwestern University, Evanston, IL 60208 \\
$^{(2)}$Fermi National Laboratory,
P.O. Box 500, Batavia, IL 60510 \\
$^{(3)}$Institute of Physics, Academia Sinica, Taipei, Taiwan, R.O.C.\\
$^{(4)}$Physics Department, University of Illinois at Chicago, IL 60680 \\
\end{instit}
%\moreauthors{C. C. Author}
%\begin{instit}
%Second author institution and/or address
%\end{instit}
%\receipt{\today}
\vskip -1cm
\begin{abstract}
We investigate a potentially large CP violating asymmetry in the neutral
Higgs boson decay into a heavy quark pair or a $W^+$--$W^-$ pair. The
source of the CP nonconservation is in the Yukawa couplings of the Higgs
boson which can contain both scalar and pseudoscalar pieces. One of the
interesting consequence is the different rates of the Higgs boson decays
into CP conjugate polarized states. The required final state
interactions can be either the strong or the electroweak interactions.
The CP violating asymmetry can manifests itself in the asymmetry in the
energies of the secondary decay products of these heavy quarks and $W$
bosons. Such asymmetry can be measureable in the future colliders such
as SSC or LHC.
\end{abstract}
\newpage
\narrowtext
\pagestyle{plain}
The physics related to the Higgs boson is the most mysterious part of
the Standard Model.  It is widely expected that whatever information we
can obtain about the Higgs sector from the next generation colliders
will give us hints about the potential new physics beyond the Standard
Model.  Most of analyses about the search for Higgs bosons, in the Standard
Model or beyond, often ignore potential CP violation. However, in these
models, the lightest neutral Higgs boson can have interesting CP
violating phenomena.  In some sense, as we shall elaborate later, the CP
violating aspect may be the most interesting part of the Higgs boson
physics beyond the Standard Model once a neutral Higgs is identified.

In this letter, we investigate an interesting signature of CP
violation in the neutral Higgs decay into a heavy quark pair $Q\bar Q$
or a $W^+W^-$ pair.
For the heavy quark $Q {\bar Q}$ mode,
the polarizations of the quarks are
either $Q_L {\bar Q}_L$ or $Q_R {\bar Q}_R$. (Note that we have adopted
the notation that ${\bar Q}_L$ is the antiparticle of $Q_R$ and should
be left handed.) These two modes happen to be CP conjugate of each
other. Therefore one can immediately consider the CP asymmetry in the
event rate difference, $N(Q_L {\bar Q}_L)-N(Q_R {\bar Q}_R)$.  For the
$W$ pair mode, the two polarization modes $W^+_L W^-_L$ and $W^+_R
W^-_R$ are also CP conjugate of each other.  (Here we denote the states
of helicities $-1$, $1$ and $0$ by $L$, $R$ and $\parallel$ respectively.)
Therefore, their difference in the decay rates, $N(W_L^+ W_L^-) - N(W_R^+
W_R^-)$, signals CP violation. For the $Z$ pair mode, one can look at
the similar asymmetry $N(Z_L Z_L) - N(Z_R Z_R)$ also.

A method in detecting the asymmetry $N(Q_L {\bar Q}_L)-N(Q_R {\bar
Q}_R)$ was proposed recently in Ref.\cite{ref:Peskin,ref:kly}. One assumes that
the quark $Q$ decays semileptonically through the usual $V-A$ weak
interaction. For a lighter quark (such as the $b$ quark), the information
about polarization is probably washed away in the soft process of
hadronization before its subsequent decay.
However, for a heavy quark such as the top quark, since the
hadronization time is much longer than the decay time\cite{ref:Bigi},
one can analyze polarization dependence of its decay at the quark level.
The top quark first decays into a $b$ quark and a $W^+$ boson,
which subsequently becomes $\bar l \nu$.
For heavy top quark, the $W^+$ boson produced in top decay is predominantly
longitudinal.
Due to the $V-A$ interaction, the $b$ quark is preferentially produced in
the left-handed helicity.  So the longitudinal $W^+$ boson is
preferentially produced along the direction of the top quark polarization.
Therefore the anti-lepton $\bar l$ produced in the $W^+$ decay is also
preferentially in that direction.
At the rest
frame of $t$, the angular distribution\cite{ref:Kuhn}
of the produced $\bar l$ has the
form $1+\cos\psi$, with $\psi$ as the angle between $\bar l$ and the
helicity axis of $t$.
When the Higgs boson decays, the top quark is produced usually with nonzero
momentum. As a result of the Lorentz boost, the anti-lepton
$\bar l$ produced in the decay of the right handed top quark $t_R$ has
a higher energy than that produced in the decay of the left handed top
quark $t_L$. Similarly, the $l$ lepton produced in the decay of ${\bar
t}_L$ has a higher energy than that produced in the decay of  ${\bar
t}_R$. Consequently, in the decay of the pair $t_L {\bar t}_L$ the
lepton from $\bar t_L$ has a higher energy than the anti-lepton
from $t_L$; while in the decay of $t_R {\bar t}_R$ the anti-lepton
has a higher energy.
Therefore one can observe $N(t_L {\bar t}_L)-N(t_R {\bar t}_R)$
by measuring the energy asymmetry in the resulting
leptons\cite{ref:OtherCP}.

For asymmetry in the $W^+ W^-$ mode, one can look at the leptonic decays
of the $W$ gauge bosons.
We are interested in the transverse W bosons in this case.
In the rest frame of $W^+$, which decays into
$\bar l\nu$, the angular distribution of $\bar l$ has the form
$(1+\cos\psi)^2$, with $\psi$ as the angle between $\bar l$ and the
helicity axis along which the spin projection of $W$ is one.
Similar to previous analysis, the anti-lepton in the decay of
$W^+_R$ has a higher energy than the lepton from the $W^-_R$, while
the lepton in the decay of
$W^-_L$ has a higher energy than the anti-lepton from the $W^+_L$.
Therefore, the relative energy between the lepton and the antilepton
in the decay of a $W^+ W^-$ pair can be translated
into the information about the polarization of $W$ bosons.
The asymmetry is much harder to detect in the leptonic final states
for the case of $Z$ pairs.
Because of the dominance of the axial-vector coupling to leptons,
there is only a weak angular dependence for the outgoing leptons.
However it is present in principle.

In order to generate the asymmetries
$N(t_L {\bar t}_L)-N(t_R {\bar t}_R)$ or
$N(W^+_L W^-_L)-N(W^+_R W^-_R)$,
it is necessary to include effects of the final state interactions
in order to escape from the hermiticity constraint at the tree level
due to the CPT theorem.
These final state interactions, as shown in Fig.~1, can come from
the strong, the electroweak, or the Higgs corrections to the $H Q {\bar
Q}$ or $H W^+ W^-$ vertices.

In the $H\rightarrow t\bar t$ channel, CP non--conservation
occurs in the complex Yukawa couplings,
\begin{equation} {\cal L}_{H\bar t t}=-(m_t/v) \bar t (AP_L+A^*P_R) t H
\;,\quad
          v=(\sqrt{2} G_F)^{-{1\over 2}} \simeq 246 \hbox{ GeV} \; .
\end{equation}
The complex coefficient $A$ is a combination of model-dependent mixing angles.
Simultaneous presence of both the real part $A_R={\rm Re}A$ and the
imaginary part $A_I={\rm Im} A$ guarantees CP asymmetry. For example, at
the low energy regime, it can give rise to the electric dipole moment of
elementary particles \cite{ref:Weinberg,ref:Barr}. Here we will show
that CP nonconservation manifests itself in the event rate difference in
collider experiments.
We denote $\delta^{\rm QCD}$, $\delta^{\gamma}$, $\delta^Z$, $\delta^H$,
$\delta^{WW}$, $\delta^{ZZ}$,  as contributions in CP asymmetry due to the
exchange of the gluon, the photon, the $Z$ boson, the Higgs bosons, or those
with intermediate states of a $W$ pair or a $Z$ pair respectively,
\begin{equation}
\Delta \equiv
      {N(t_L \bar t_L)-N(t_R \bar t_R)
\over  N (\hbox{all } t\bar t \hbox{ from } H)}
={\delta^{\rm QCD}+\delta^{\gamma}+\delta^{Z}+\delta^{H}
 +\delta^{WW} +\delta^{ZZ}
                             \over
                    \beta_t^2 A^2_R +A^2_I} ,
\label{eq:asympol}
\end{equation}
with $\beta_t^2=1-4m_t^2/M_H^2$.
The one--gluon exchange gives large CP asymmetry,
\begin{equation}  \delta^{\rm QCD} = (C \alpha_S /2){\rm Im}(A^2) (1-\beta_t^2)
 \;,
\end{equation}
with the color factor $C=4/3$.
It is interesting to note that the imaginary part of the one--loop graph
contributes a factor of $A_R$ while the tree
graph contributes $A_I$ (that is, the pseudoscalar coupling).
When $A_I \sim A_R$, the asymmetry is of order
of the strong coupling $\alpha_S$, about $10^{-1}$.
Fig.~2 shows how such asymmetry depends on $M_H$.
Note that there is no strong constraints\cite{ref:Weinberg,ref:Barr} on
Im$(A^2)$, which can easily be of order one.

The electromagnetic correction $\delta^\gamma$
is obtained by replacing $\alpha_S$ by the QED coupling $\alpha$,  and
$C$ by 4/9, the charge squared of the $t$--quark. The result is
negligible.
The contribution by the $Z$--exchange is
\begin{equation} \delta^Z ={\sqrt{2} G_F m_t^2\over 8\pi} {\rm Im}(A^2) \Bigl\{
        [F({M_H^2\beta_t^2\over M_Z^2})-2]\beta_t^2 +
     {M_Z^2 \over M_H^2 }F({M_H^2\beta_t^2\over M_Z^2})[1+(1-8x_W/3)^2] \Bigr\}
\;.
\end{equation}
Here the function $ F(x) = 1-x^{-1}\log(1+x)$,
and $x_W=\sin^2\theta_W\approx 0.23$.
The term with the factor $(1-8x_W/3)^2$ is due to the
part of the one--loop graph in which both vertices of $Z$ boson are
vectorial while the rest are due to the part in which both vertices of
$Z$ are axial--vectorial. The parts with one axial vector vertex and one vector
vertex do not contribute.  For the part involves only the axial vector
vertices of $Z$ boson, the tree graph can contributes either scalar or
pseudoscalar coupling.
Contribution from the Higgs bosons is more complicated.
In general, there are more than one neutral Higgs bosons $H_i$,
with Yukawa couplings to the $t$--quark of the form,
${\cal L}_{H_i\bar t t}= - (m_t/v) \sum_i \bar t (a_i P_L +a_i^* P_R) t H_i$.
The coefficient $A$ is only one of these $a_i$.
The overall effect from all Higgs bosons is
\begin{equation} \delta^H = {\sqrt{2} G_F m_t^2 \over 8 \pi} \sum_i
    F({M_H^2\beta_t^2\over M_{H_i}^2})
                     \Bigl[
    \beta_t^2 {\rm Im}(A^2 a_i^{*2})
  + (1-\beta_t^2) {\rm Im}(A) {\rm Re}(A |a_i|^2 + A a_i^{*2})
                                    \Bigr].
\end{equation}
This general formula is valid for the decay of each Higgs boson in the
system. For the purpose of illustration, we work in a simplified case of
the decay of the lightest Higgs boson $H$. Then, if the effects of
heavier Higgs bosons can be neglected, the formula is much reduced as
$a_H=A$ and the first term in the bracket of the above equation
vanishes. Numerical study shows that $\delta^H$ is very small in this
scenario.

There are important contributions from processes involving $W^+W^-$
or $ZZ$ intermediate states as shown in Fig.~1. The amplitudes depend
on the the vertices,
\begin{equation}{\cal L}_{HWW,HZZ}=B g_2 H [ M_W W^{+\nu} W^-_\nu
                       +\case 1/2 (M_Z/\cos\theta_W) Z^\nu    Z_\nu] \; .
\end{equation}
In any model, the scalar boson that couples to the $W$ pair or $Z$ pair
is the scalar partner of the unphysical Higgs boson.  This scalar boson
is in general not a mass eigenstate.
Therefore we parametrize its coupling by a phenomenolgical coefficient $B$
in addition to the usual $SU_L(2)$ gauge coupling $g_2$.
The coefficient $B$ represents the mixing factor needed to reach the mass
eigenstate.   It reduces to unity in the Standard Model
with one single Higgs doublet.
For the diagrams with $W^+W^-$ intermediate states,
we obtain the $CP$ asymmetry,
\begin{eqnarray}
\delta^{WW}=-{\sqrt{2}G_FM_W^2 BA_I \beta_W\beta_t\over 4\pi}
&\Bigl\{&
   \Bigl[ {3\beta_t^2-2\beta_W^2-1 \over 2\beta_t^2}
       +{1-\beta^2_t\over \beta_t^2(1-\beta^2_W)}
   \Bigr] L(\beta_W,\beta_t)   \nonumber\\
&\;&
   -2L(\beta_W,\beta_t)+2+{1+\beta_W^2\over 1-\beta^2_W}
\Bigr\}               \;,
\end{eqnarray}
where  $\beta_W^2=1-4M_W^2/M_H^2$, and the function
\begin{equation}  L(x,y) \equiv
           1+{x^2-y^2\over 2xy}
       \log \Bigl\vert {x-y \over x+y} \Bigr\vert \;.
\end{equation}
Also, for the diagrams with $ZZ$ intermediate states,
we obtain the $CP$ asymmetry,
\begin{eqnarray}
\delta^{ZZ}=-{\sqrt{2} G_F M_Z^2 BA_I\beta_Z\over 16\pi\beta_t}
   &\Bigl[&
     (1-{8\over3}x_W)^2 ( 1+{1
      +\beta_Z^2+2\beta_t^2\over 4\beta_Z\beta_t} {K})(1-\beta_Z^2)
                 \nonumber\\
   &\;& + (1-\beta^2_Z)(1+{1+\beta_Z^2-2\beta_t^2\over 4\beta_Z\beta_t} {K})
                                                 \\
   &\;&       + {1-\beta_t^2\over 1-\beta_Z^2}
                  {(1+\beta_Z^2)^2 \over 2\beta_Z\beta_t} K
                              +2{1+\beta_Z^2\over 1-\beta_Z^2} \Bigr]
   \;,             \nonumber
\end{eqnarray}
with
\begin{equation} K=\log\Bigl\vert {1+\beta_Z^2-2\beta_t\beta_Z
                        \over 1+\beta_Z^2+2\beta_t\beta_Z}
     \Bigr\vert  \;.
\end{equation}
The first term in the square bracket is the contribution where both of
the $Z$ couplings with the top quark are vectorial.  The remaining two
terms correspond to the contribution where both couplings are
axial-vectorial.
Also note that the one--loop induced vertex is always scalar and
the $A_I$ factor arises from the tree level amplitude.
Numerical study shows that $\delta^{WW}$ and $\delta^{ZZ}$ are very important
for the natural scenario $A_R \sim B$. Their contributions can dominate  when
$M_H$ is large as illustrated in Fig.~2.
In this letter, we shall ignore the contributions
due to tri-Higgs couplings because they are more model dependent in general.
Also, this type of contributions disappears,
due to the vanishing imaginary part, when the decaying Higgs
boson is the lightest Higgs boson.
In any case, this type of contribution will not affect our results very
much as long as no accidental cancellation occurs.

Since CP violation in Eq.(1) would disappear if the top quark were massless,
one may wonder why the above formulas for $\delta^{ZZ}$ and
$\delta^{WW}$ are not proportional to $m_t$.  The answer is that, in the
denomenator, the leading (tree) level contribution also has a factor of
$m_t$ is our parametrization in Eq.(1).  The $m_t$ required physically
in the numerator is artificially cancelled by this factor in the
denomenator.

The polarization asymmetry is Eq.(\ref{eq:asympol}) can be translated in
to the lepton energy asymmetry\cite{ref:Peskin,ref:kly,ref:ckp}. The
energy $E_0(l^+)$ distribution of a static $t$ quark decay $t\rightarrow
l^+\nu b$ is very simple\cite{ref:Kuhn} in the narrow width $\Gamma_W$
approximation when $m_b$ is negligible.
\begin{equation}
    f(x_0)= \left\{
              \begin{array}{lll}
       x_0 (1-x_0)/D & \quad\  & \mbox{if $m_W^2/m_t^2 < x < 1$}, \\
       0             & \quad\  & \mbox{otherwise.}
             \end{array}
              \right.
\end{equation}
Here we denote the scaling variable $x_0=2E_0(l^+)/m_t$ and
the normalization factor $D={1\over 6}-{1\over 2}(m_W/m_t)^4
+{1\over 3}(m_W/m_t)^6$.
When the $t$ quark is not static, but moves at a speed $\beta$
with helicity $L$ or $R$, the distribution expression becomes
a convolution,
\begin{equation}
     f_{R,L}(x,\beta)=
    \int_{x/(1+\beta)}^{x/(1-\beta)} f(x_0)
            {\beta x_0 \pm (x-x_0) \over 2 x_0^2\beta^2}
      dx_0
\;.
\end{equation}
Here $x=2E(l^+)/E_t$. The kernel above is related to the
$(1\pm\cos\psi)$  distribution mentioned in the introduction.
Similar distributions for the $\bar t$ decay is related by CP
conjugation at the tree--level.
Using the polarization asymmetry formula in  Eq.(\ref{eq:asympol}),
we can derive expressions for the energy distributions of
$l^-$ and $l^+$:
\begin{equation}
N^{-1}dN/dx(l^\pm)
=\case 1/2(1\pm\Delta)f_L(x,\beta_t)+\case 1/2(1\mp\Delta)f_R(x,\beta_t)
\;.
\label{eq:dist}
\end{equation}
Here distributions are compared at the same energy for the lepton
and the anti--lepton at the $H$ rest frame,
$x(l^-)=x(l^+)=x=4E(l^\pm)/M_H$.
To prepare a large sample for analysis, we only require that each event
has at least one prompt anti--lepton $l^+$
from the $t$ decay {\it or}
one prompt lepton $l^-$ from the $\bar t$ decay.
Fig.~3 compares the lepton and anti--lepton energy distribtuions for a
typical asymmetry $\Delta=0.1$.
One can also sum over channels of different lepton flavor to
increase the event rate.

For the $W$ pair production, the tree level amplitude,
parametrized by $B$, interferes with the fermion--loop amplitude
(Fig.~4) of the $b$ quark exchange to
produce the CP violating asymmetry.
The resulting asymmetry is
\begin{equation}
\Delta_W \equiv
      {N(W^+_L W^-_L)-N(W^+_R W^-_R)
\over  N(\hbox{all } W^+ W^-,\hbox{ from }H) }
= {3\sqrt{2} G_F m_t^2 \over 4\pi B }A_I{\beta_t\over \beta_W}
    L(\beta_t,\beta_W)
\Bigl({2(1-\beta_W^2)^2\over 3-2\beta_W^2+3\beta_W^4} \Bigr)
       \;.
\label{eq:ww}
\end{equation}
Note that, for a heavy Higgs boson, its leading mode is neither $W^+_L
W^-_L$ nor $W^+_R W^-_R$, but the longitudinally polarized state of
$W^+_\parallel W^-_\parallel$.
However, the $W^+_\parallel W^-_\parallel$ state is CP self conjugate
and cannot provide the CP asymmetry we are looking for.
Consequently, there is a suppression factor in the last bracket of
Eq.({\ref{eq:ww}}).
The factor is the ratio between the transverse $W^-W^+$ rate
and the overall $W^-W^+$ rate from the Higgs boson decay.
Fig.~5 shows such asymmetry, which is sizable near the threshold of
$t\bar t$.
As expected, when the $t\bar t$ threshold gets higher the asymmetry gets
smaller due to the suppression of the transverse modes mentioned above.
This asymmetry can be detected by measuring the
energy asymmetry of the oppositely charged leptons $l^\pm$ in the decays
of the $W^\pm$ bosons.
We define the dimensionless variable $x(l)=4E(l)/M_H$ in the $H$ rest
frame as before. The $x$ distributions due to the process
$H\rightarrow  W^+W^-$ are given by
\begin{eqnarray}
{1\over N}{dN\over dx(l^\pm)}
&=&
\Bigl({(1+\beta_W)^2\over 3-2\beta_W^2+3\beta_W^4}\Bigr)
                      {3[\beta_W^2-(1-x)^2] \over 4 \beta_W^3}
                                       \nonumber\\
&+&
\sum_{s=-1,+1}
\Bigl({(1-\beta_W^2)^2\over 3-2\beta_W^2+3\beta_W^4}
- {s\Delta_W\over 2} \Bigr)
{3(x-1\pm s\beta_W)^2\over 8\beta_W^3}
\; .
\end{eqnarray}
The first term above comes from the longitudinal $W$, and
the rest are due to the two transversely polarized $W$'s
(helicity $s=-1,+1$). Each bracket corresponds to the production
weight of the corresponding polarized state.
Fig.~6 compares the lepton and anti--lepton energy distribtuions for a
typical asymmetry $\Delta_W=0.01$.
As before, we only require that each event
has at least one prompt anti--lepton $l^+$
from the $W^+$ decay {\it or}
one prompt lepton $l^-$ from the $W^-$ decay.
It is worthwhile to point out that
the smallness of $\Delta_W$ is mostly due to
the large contribution of the longitudinal $W$ mode.
For example, this suppression factor, in the last bracket of
Eq.(\ref{eq:ww}),
is $0.015$ when $M_H = 400$ GeV.
In Fig.~6,
We expect that events with the lepton energy $x \sim (1\pm\beta_W)$ come
from the transverse modes, and otherwise, the central bump is dominated
by the longitudinal $W$ mode.
Therefore, in principle, one can
enhance effects from the transverse modes
by imposing energy cuts that eliminate events in the central region from
the longitudinal mode. The efficiency of such cuts deserves detailed
simulations in the future\cite{ref:Long}.

Similar CP asymmetry can be formed between the $Z_L Z_L$ and
the $Z_R Z_R$ pairs,
\begin{equation}
 {N(Z_L Z_L)-N(Z_R Z_R) \over  N(Z_L Z_L)+N(Z_R Z_R) }
= {3\sqrt{2} G_F m_t^2 \over 4\pi B }A_I{\beta_t\over \beta_Z}
    \Bigl[1+{K\over 4\beta_t\beta_Z}-
        (1-{8\over 3}x_W)^2 {K\beta_Z\over 4\beta_t} \Bigr]
       \;.
\end{equation}
However it is not clear how feasible one can decode such
asymmetry from the final decay products of the $Z$'s.

In conclusion, we have shown a potentially large CP violating asymmetry
in the decay of the neutral Higgs boson.
For the channels $H\rightarrow t\bar t$, $W^+W^-$, $ZZ$, to be
available, the Higgs mass has to be about $180$ GeV or higher.
That rules out LEP~II and possibly Tevatron as the production machine.
The heavy Higgs boson can be produced copiously
in the future hadron
collider via the gluon-gluon or $W$-$W$ fusions, or in the next
high energy $e^+e^-$ machine via $Z$-bremsstrahlung or $W$-$W$ fusion.
At SSC, if $200\hbox{GeV} \leq M_H \leq 1 \hbox{TeV}$,
there will be between $10^6$ and $10^5$ Higgs boson
produced\cite{ref:haber} for an annual integrated luminsity of $10^4
\hbox{pb}^{-1}$.
The ``gold-plated'' events, $H \rightarrow Z Z\rightarrow
(l^+l^-)(l^+l^-)$, are supposed to be the discovery mode for the Higgs
mass up to about $600$ GeV.  The $W^+ W^-$ mode is about twice as
prolific as the $ZZ$ mode and has even larger leptonic branching
fraction. However, their signature of a charged lepton plus two jets
from the $W$ has large non--resonant background in the Standard Model.
More works\cite{ref:Long} are needed in decoding the signature.
The branching fraction for the Higgs boson decay into $t \bar{t}$ can be
between $5\%$ and $20 \%$ for
$350 \hbox{GeV} \leq M_H \leq 1 \hbox{TeV}$ and $m_t = 150 \hbox{GeV}$
for example.
Their signature is just as hard to decode as that of the $W^+W^-$ mode.
However, once the Higgs mass is known from the $ZZ$ mode, the $W^+W^-$
and $t\bar t$ modes should be much easier to identify.

D.C. wishes to thank the Institute of Physics at Academia Sinica for
hospitality
while this work was being completed.
We also wish to thank I. Phillips, K. Cheung and M. Peskin for discussion.
This work is supported by grants from
Department of Energy and from National Science Council of Republic of China.

\figure{Feynman diagrams for the  process $H\rightarrow t\bar t$.
The tree amplitude (a) interferes with those one--loop amplitudes with
the final state interactions coming from
(b) the exhange of a gluon $g$, a photon $\gamma$,
a $Z$ gauge boson or a Higgs boson $H_i$;
(c) the intermediate $W^+W^-$ boson pair;
(d) the intermediate $ZZ$ boson pair.
}
\figure{
$[N(t_L \bar t_L) - N(t_R \bar t_R)]
/ N(\hbox{all } t\bar t, \hbox{ from }H)$
versus $M_H$  for $m_t=150$ GeV.
The illustration is for the case $A_I=A_R=B=0.5$.
}
\figure{ The energy distribution of the anti--lepton (lepton) in the $H$
rest frame is plotted in the solid (dashed) curve, for the case that
$H\rightarrow t\bar t$, $M_H=300$ GeV, $m_t=120$ GeV, and $\Delta=0.1$.
The asymmetry is shown in the bottom.}
\figure{(a) The one--loop diagram for $H\rightarrow W^+W^-$ produces
the CP violating asymmetry when it interferes with the tree amplitude.
(b) The same effect occurs for $H\rightarrow ZZ$.
}
\figure{
$[N(W_L^+W_L^-)-N(W_R^+W_R^-)]
/ N(W^+W^-, \hbox{ from }H)$
versus $M_H$ for different values of
$m_t$=100 GeV (solid), 150 GeV (dashed),
200 GeV (dotted), and 250 GeV (dash--dotted).}
\figure{ Energy distributions of the lepton and the anti--lepton in the
$H$ rest frame are compared, for the case that $H\rightarrow W^+W^-$,
$M_H=300$ GeV, and $\Delta_W=0.01$.
The asymmetry is shown in the bottom.}
\end{document}